\documentstyle[12pt]{article}
\def\beq{\begin{equation}}
\def\eeq{\end{equation}}
\def\bea{\begin{eqnarray}}
\def\eea{\end{eqnarray}}
\def\bq{\begin{quote}}
\def\eq{\end{quote}}
\def\ve{\vert}
\def\nnb{\nonumber}
\def\ga{\left(}
\def\dr{\right)}
\def\aga{\left\{}
\def\adr{\right\}}

\def\rar{\rightarrow}
\def\nnb{\nonumber}

\def\nin{\noindent}
\begin{document}
\topmargin -2.0cm
\oddsidemargin +0.2cm
\evensidemargin -1.0cm
\pagestyle{empty}
\begin{flushright}
{CERN-TH.7166/94}\\
PM 94/06
\end{flushright}
\vspace*{5mm}
\begin{center}
\section*{
 \boldmath{$B \rar K^* \ \gamma$} FROM
HYBRID SUM RULE }
\vspace*{1.5cm}
{\bf S. Narison} \\
\vspace{0.3cm}
Theoretical Physics Division, CERN\\
CH - 1211 Geneva 23\\
and\\
Laboratoire de Physique Math\'ematique\\
Universit\'e de Montpellier II\\
Place Eug\`ene Bataillon\\
34095 - Montpellier Cedex 05\\
\vspace*{2.0cm}
{\bf Abstract} \\ \end{center}
\vspace*{2mm}
\noindent
Using the {\it
hybrid} moments-Laplace sum rule (HSR), which
is well-defined for $M_b \rar \infty$, in contrast with
the $popular$ double Borel (Laplace) sum rule (DLSR),
 which blows up in this limit when applied to
  the heavy-to-light processes,
we
show that the form factor of the $B \rar K^* \  \gamma$
radiative
transition
is dominated
by the light-quark condensate for $M_b \rar \infty$ and behaves like
$\sqrt M_b$.
The form factor
is found to be
$
F^{B\rar K^*}_1(0) \simeq (30.8 \pm 1.3
\pm 3.6 \pm 0.6)\times 10^{-2},
$
where the errors come respectively from the procedure in the sum rule
analysis, the errors in the
input and in the $SU(3)_f$-breaking parameters.
This result leads to $Br(B\rar K^* \ \gamma) \simeq (4.45 \pm 1.12)
\times 10^{-5}$ in agreement with the recent CLEO data.
Parametrization of the
$M_b$-dependence of the form factor
including the  $SU(3)_f$-breaking effects
is given in (26), which leads to
$F^{B\rar K^*}_1(0)/ F^{B\rar \rho}_1(0) \simeq (1.14 \pm 0.02)$.

\vspace*{2.0cm}
\noindent


\begin{flushleft}
CERN-TH.7166/94 \\
PM 94/06\\
February 94
\end{flushleft}
\vfill\eject
\pagestyle{empty}

\setcounter{page}{1}
\pagestyle{plain}

\section{Introduction}
With the advent of the heavy-quark symmetry \cite{MAN},
there has been
considerable interest and progress in the understanding of the
semileptonic
form factors of the transition of a heavy quark
into another heavy quark, since
 in this infinite mass limit
all semileptonic form factors reduce to the single Isgur-Wise
function \cite{IWI}.
In the case of heavy-to-light transitions, Isgur  and Wise \cite{IWI2}
have remarked that, if one assumes that only the upper two components
of the $b$
 quark field contribute, there will be relations between various
form factors of the semileptonic and rare $B$-decays. The relations
obtained from the static heavy quark approach are not rigorous
as the momentum transfer of e.g, the $B \rar K^* \gamma$, which is of the
order of $M_b/2$, is large, in such a way that perturbative contributions
related to the hard process \cite{BROD}
can invalidate the Isgur Wise relations. Burdman and
Donoghue \cite{DON} have shown that in these processes,
the soft quark model contributions dominate over the perturbative ones,
such that the results of Isgur-Wise might still
be valid and could be extended to the whole range
of $q^2$.

\nin
In this paper, we shall examine the validity of
the previous results, using the approach
of QCD Spectral Sum Rules (QSSR) in the analysis of the
form factor for the
$B \rar K^* \gamma $ radiative process. The form factor is
 defined as:
\bea
<K^*(p')\ve \bar s \sigma_{\mu \nu}\ga
\frac{1+\gamma_5}{2}\dr q^\nu b\ve B(p)> =
i\epsilon_{\mu \nu \rho \sigma}\epsilon^{*\nu}p^\rho p'^\sigma
F^{B\rar K^*}_1
\nnb \\
+\aga \epsilon^*_\mu(M^2_B-M^2_{K^*})-\epsilon^*q(p+p')_{\mu}
\adr \frac{F^{B\rar K^*}_1}{2}.
\eea
In the QSSR evaluation of the form factor , we shall consider the
generic three-point function
(omitting Lorentz indices):
\beq
V(p,p',q) = -\int d^4x \int d^4y\, \mbox{exp}
(ip'x-ipy) <0\ve TJ_q(x)O(0)J_b(y)\ve
0>,
\eeq
whose Lorentz decompositions are analogous to the previous hadronic
amplitudes; $J_q \equiv \bar s \gamma^\mu d
$ is the bilinear quark current having the quantum
numbers of the  $K^*$; $J_b \equiv (M_b +m_d)
\bar d (i\gamma_5)b$ is
the quark current associated to the $B$-meson;
 $  O\equiv
\bar b \frac{1}{2}\sigma_{\mu \nu}q^{\nu}s $ is the weak operator.
The vertex function obeys the double dispersion relation:
\beq
V(p^2,p'^2,q^2)= \frac{1}{\pi^2}\int_{M^2_b}^{\infty}\frac{ds}{s-p^2}
\int_{0}^{\infty}\frac{ ds'}{s'-p'^2} \,
\mbox{Im}\, V(s,s',q^2),
\eeq
where $q \equiv p-p'$ is the momentum transfer.
\section{Choice of the sum rule}
Widely used in the literature is the double exponential Laplace (Borel)
sum rule (DLSR), which
has given successful predictions in the heavy-to-heavy transitions.
This method has been used also for the heavy-to-light transitions
\cite{DOS}--\cite{COL}. Though at finite $b$-quark mass value,
the numerical fits of the form factors may be quite good, one can
notice
 that the OPE in the DLSR blows up
for $M_b \rar \infty$, which invalidates the uses of the DLSR in this
channel.
 This is due to the existence of terms of the
type:
\beq
\frac{M^{2l}_b}{\ga M^2_b-p^2 \dr^{k} p'^{2k'}},
\eeq
which appear
in the successive evaluation of the Wilson coefficients of high-dimension
operators.
After the double Laplace (Borel) transform,
 these terms convert into:
\beq
{M^{2l}_b}\tau^{k}\tau'^{k'} \mbox{exp}(-\tau M_b^2),
\eeq
where $\tau$ (resp. $\tau'$) is the sum-rule variable associated
to the heavy (resp. light) quark.
In this limit, it is convenient to introduce the non-relativistic
sum-rule variable $\tau_{NR} \equiv M_b \tau$, where $\tau_{NR}$ is
$M_b$-independent. Then, it is clear from (5) that
the OPE converges {\it if and only if} $k \geq 2l$. This
condition is not fulfilled in the case of the
$B \rar K^* \ \gamma$ process,
as also
noticed by \cite{ALI}, \cite{COL}, while we have also checked that it
is not satisfied in some of the form factors of the $B \rar \rho$
and $D \rar K^*$
semileptonic processes analyzed in \cite{DOS},\cite{BAL}, though this
does not effect the numerical estimates of these form factors.
In order to
restore the good behaviour of the OPE,
the authors of Ref.
\cite{COL} modelize the condensates in an {\it ad hoc} way.
We do not find
this argument convincing. Moreover, this parametrization
cannot guarantee the
convergence of the OPE at higher orders in dimensions.

\nin
An alternative promising direction
is to find a sum rule, other than DLSR, which is
more appropriate for this channel.
The
light-cone sum rule has been chosen in
Ref. \cite{ALI};
 but, due to the complexity (and, perhaps, less understood structure)
of the wave functions entering in
this sum rule, we do not have here the simplicity
and the transparency of the original
SVZ-expansion, where, in this case, the non-perturbative
effects are simulated by the vacuum condensates.

\nin
In this paper, we shall present
another type of sum rule, which is in the
line of the SVZ-expansion and which
is adequate for the heavy-to-light process, since it has a good behaviour
 when
$M_Q \rar \infty$. This sum
rule  has been
invented in \cite{SN1} for the
analysis of the $B \rar D, \ \pi $ and $
\rho$ semileptonic decays. This sum rule
combines the moment (finite number of derivatives evaluated at $p^2=0$)
which is good
for the heavy quark as $M_b \gg \Lambda$ (1/$M_b$-expansion),
and the Laplace one, which is
appropriate
for the light quark if $p'^2 \gg \Lambda^2$ (1/$p'^2$-expansion).
 We shall hereafter call
this sum rule, as in \cite{SN1}, the
{\it hybrid sum rule} (HSR). It has the form:
\bea
\cal {H}(n,\tau') &\equiv &
\frac{1}{n!}\ga\frac{\partial}{\partial p^2}\dr^n
_{p^2=0}\cal{L}\ga V(p^2,p'^2,q^2) \dr  \nnb \\
&=&
\frac{1}{\pi^2}\int_{M^2_b}^{\infty}\frac{ds}{s^{n+1}}
\int_{0}^{\infty} ds' \, \mbox{exp}(-\tau' s')
\mbox{Im}\, V(s,s',q^2),
\eea
where $\cal{L}$ is the Laplace (Borel) exponential operator. As one can
obviously  notice, the {\it hybrid} transform of (4) is:
\beq
\frac{\tau'^{k'}}{ M_b^{2(k+n-l)}},
\eeq
 which shows that the convergence of the OPE for $M_b \rar \infty$ is
reached with the much weaker condition $k \geq l-n$
than the one $k\geq 2l$
of the DLSR. Indeed, in the specific case of the $B \rar
K^* \ \gamma $ process, the truncated series with the inclusion of the
mixed condensate already
converges for
$n= 0$.

\nin
In order to come to observables, we
insert intermediate states between the electromagnetic
and hadronic currents in (2),
while we smear the higher-states effects with
 the discontinuity of the QCD
graphs from a threshold $t_c$ ($t'_c$) for the heavy (light)
mesons. Therefore, we have the sum rule for the form factor
$F^{B\rar V}_1(q^2)$:
\bea
\cal{H}_{res}& \simeq& 2C_V f_B \frac{F^{B\rar V}_1(q^2)}{M_B^{2n}}
\mbox{exp} \, (-M^2_V\tau)
\nnb \\
&\simeq&
 \frac{1}{\pi^2}\int_{M^2_b}^{t_c}\frac{ds}{s^{n+1}}
\int_{0}^{t'_c} ds' \, \mbox{exp}(-\tau s')
\mbox{Im}\, V_{PT}(s,s',q^2) + \mbox{NPT}.
\eea
PT (NPT) refers to perturbative (non perturbative)
contributions;
$C_{V}\equiv M_{V}^2/(2\gamma_V)$
for light vector mesons; $M_V$ is the
light-meson mass. The
decay constants are normalized as:
\bea
 (m_q+M_Q)<0\ve \bar q (i\gamma_5)Q\ve P>= \sqrt 2 M^2_Pf_P \nnb \\
<0\ve \bar q \gamma_\mu Q\ve V> =\epsilon^*_\mu \sqrt 2 \frac{M^2_V}
{2 \gamma_V}.
\eea
\section{ HSR estimate of the
\boldmath{$B \rar \rho \ \gamma$} form factor }
In the following, we shall use the previous sum rule for the estimate
of the $B \rar \rho \ \gamma$ form factor.
The QCD expression
 of the corresponding vertex function reads:
\bea
\mbox{Im} V(s,s')&=&\frac{3}{8} \frac{s'M^5_b}
{(s-s')^3} \nnb \\
 V_{ qq}&=&-\frac{M^2_b}{2}\frac{<\bar dd>} {(M^2_b-p^2)(-p'^2)}.
\eea
We shall use the contribution of the mixed condensate obtained by
\cite{ALI}. Then, we deduce the sum rule:
\bea
\cal{H}_{res}
&\simeq&
 \frac{3}{8\pi^2}M_b^5\int_{M^2_b}^{t_c}\frac{ds}{s^{n+1}}
\int_{0}^{t'_c} ds'\, \frac{s'}{(s-s')^3} \, \mbox{exp}(-\tau s')
\nnb \\
&-& \frac{ <\bar dd>(\tau')}{2M_b^{2n}} \aga
1- {\tau' M_0^2}\ga \frac{(n+1)}{3}
+ \frac{\tau'^{-1}}{4 M_b^2}\ga
n^2+3n+4 \dr \dr \adr.
\eea
We shall also introduce an analogous expression of the decay constant
$f_B$ from moments sum rule at the same order \cite{SN2}:
\beq
\frac{2f_B^2}{{(M_B^2)}^{n-1}}
\simeq
\frac{3}{8\pi^2}M_b^2 \int_{M^2_b}^{t_c}
\frac{ds}{s^{n+1}} \; \frac{(s-M_b^2)^2}{s}
-\frac{<\bar dd>}{M_b^{2n-1}}\aga 1-\frac{n(n+1)}{4} \;
\ga \frac{M_0^2}{M_b^2}\dr\adr .
\eeq
For convenience, we shall work with the
non-relativistic energy parameters $E$ and $\delta M_{(b)}$:
\beq
s \equiv (M_b+E)^2, \; \; \; \;
\delta M_{(b)} \equiv M_B-M_b,
\eeq
where, as we have seen in the analysis of the two-point correlator,
the continuum energy $E_c$ is
\cite{SN2}, \cite{ZAL}:
\bea
E^D_c &\simeq& (1.08 \pm 0.26)\; \mbox{GeV}, \nnb \\
E^B_c &\simeq& (1.30 \pm 0.10) \; \mbox{GeV}, \nnb \\
E^{\infty}_c &\simeq& (1.50 \sim 1.70) \; \mbox{GeV}.
\eea
 Using (8), (10)--(14), we calculate the
form factor $F^B_1(0)$ in Fig. 1 for different values of n, $\tau'$,
$E_c$ and $t'_c$.
We use the following values of the QCD parameters for 5 flavours
\cite{SN3,SN5}:
\bea
M_0^2 &=&(0.8 \pm 0.1) \; \mbox{GeV}^2, \; \; \; \; \; \;
\Lambda \;= \; (175 \pm 41) \; \mbox{MeV}, \nnb \\
M_b &=& (4.59 \pm 0.05) \; \mbox{GeV}, \; \; \; \;
M_c = (1.47 \pm 0.05) \; \mbox{GeV}, \nnb \\
<\bar dd>(\tau')
 &=& -(189 \;\mbox{MeV})^3 \ga -\log{\tau'^{1/2}\Lambda}\dr ^{12/23},
\eea
and the experimental value $\gamma_{\rho} =
(2.55\pm 0.06)$ from the $\rho$-meson electronic width.
  Values of $\tau'$ and $t'_c$ at which this
experimental number is reproduced from the $\rho$
sum rule are \cite{SN3}:
$\tau' \simeq (0.6 \sim 1.0) \, \mbox{GeV}^{-2}$ and $t'_c \simeq
(1.7 \pm
0.3) \,\mbox{GeV}^2$. As can be noticed from Fig. 1, the values of $\tau'
$ corresponding to the stability are consistent with the previous ones.
Increasing values of $n$
 tend to destroy the existence of the stability points
due to the increase of the anomalously large
values of the $1/M^2_b$ contributions. The inclusion of higher-dimension
condensates in the OPE should restore the stability for larger values
of $n$. In our present truncated series,
the different  effects remain still corrections to the $<\bar qq>$
 condensate ones,
for $n\leq 2$.
Moving the values of $E^B_c$ and $t'_c$
within the previous ranges modifies
the shape of the curves but affects only slightly the value
of the stability point; $\Lambda$ and $M_b$
introduce each an error of 4 and 2\%.
We do not expect that radiative
corrections will affect this result in a sensible way from different
 experiences
of calculating similar observables in the heavy-to-heavy transition form
factors. Indeed, large radiative corrections
due to the coulombic-like interactions
cancel out in the ratio of the three- over the two-point function sum
rules while the radiative corrections due
to the light quark condensate is known to be
small in the cases of heavy and of light quark processes. In these
cases, the
total effect due to the radiative corrections is about 4-5\%,
which we consider as another source of errors.
Taking into account these different
sources of errors, we obtain from the
HSR:
\beq
F^{B\rar \rho}_1(0) \simeq  (27.0 \pm 1.1\pm 3.2)\times 10^{-2},
\eeq
where the first error comes from the sum-rule procedure and the
second one from the input parameters.
Extending this analysis to the $D$ mass,
we get,
for the hypothetical $D\rar \rho \ \gamma$ process:
\beq
F^{D\rar \rho}_1(0) \simeq  (62.0 \pm 10.0)\times 10^{-2}.
\eeq
\section{\boldmath{$M_b$}-dependence of the
\boldmath{$B \rar \rho \ \gamma$} form factor}
In order to understand the meaning of the previous results, let us study
{\it analytically} the sum rule
at large values of $M_b $. Since we shall work with the full theory
of QCD, the pseudoscalar quark current associated to
the $B$ meson does not acquire any anomalous dimension.
Using \cite{SN2}
\bea
f^2_B &\simeq & \frac{1}{\pi^2}
\frac{\ga E^B_c \dr ^3}{M_b}\ga \frac{M_B}{M_b} \dr^{2n-1}
\Bigg\{ (1-\frac{3}{2}(n+1)\ga
\frac{E^B_c}{M_b}\dr +\frac{3}{5}\ga (2n+3)(n+1)+
\frac{1}{4} \dr \ga \frac{E^B_c}{M_b} \dr ^2 \nnb \\
& -&\frac{\pi^2}{2}\frac{<\bar dd>}{\ga E^B_c\dr ^3} \ga 1-
\frac{n(n+1)}{4}
\frac{M_0^2}{M^2_b}\dr
\Bigg\},
\eea
from (12), we can deduce from (11):
\beq
F^{B\rar \rho}_1(0)
 \simeq  -\frac{\pi}{4}\frac{ \sqrt{M_b}}{\ga E^B_c\dr ^{3/2}}
 \, \frac{<\bar dd>}{C_{\rho}}
\mbox{exp}(M^2_{\rho}\tau')
\ga 1+ \delta^{(0)}+ \frac{\delta^{(1)}}{M_b}
 + \frac{\delta^{(2)}}{M^2_b}  \dr,
\eeq
with:
\bea
\delta^{(0)} &=&  -\frac{(n+1)}{6}M_0^2\tau' + \frac{\pi^2}{4}
\frac{<\bar qq>}{\ga E^B_c\dr ^3} \nnb \\
\delta^{(1)}& = & \frac{3}{4}(n+1)E^B_c +
\ga n+\frac{1}{2}\dr \delta M_{(b)}
\nnb \\
\delta^{(2)}
& =& -  \frac{2\cal{I}}{<\bar dd>}
 -\frac{M_0^2}{4}\ga 1+n(n+1)\frac{\pi^2}{4}\frac{<\bar qq>}
 {\ga E^B_c\dr ^3}\dr
+\frac{3\ga E^B_c \dr ^2}{640}(83n^2+230n+163),
\nnb \\
\eea
where:
\beq
\cal{I} \equiv
\frac{3}{8\pi^2}\int_{0}^{E^B_c}
\frac{2dE}{\ga 1+\frac{E^B_c}{M_b} \dr^{2n+1}
}\int_{0}^{t'_c} ds'\, \frac{s'}
{\ga \ga 1+\frac{E}{M_b}\dr^2-s'\dr^3}
\, \mbox{exp}(-\tau s').
\eeq
We have checked that this approximate expression gives a
slightly lower (about $20\%$) value of $F^{B\rar \rho}_1(0)$.

\noindent
We evaluate {\it numerically}
the coefficients of the $1/M_b$ and $1/M^2_b$ terms
 at the values $n \simeq 0\sim 1$ and $\tau'\simeq 0.5\sim 0.7
 $ GeV$^{-2}$,  where the HSR
optimizes,
from the full non-expanded expression of the three-point
function.
We use the expression of $f_B$ given by (18), which naturally has
the expected large $M_b$ behaviour.
 Then, we
deduce the
interpolating formula in units of GeV:
\beq
F^{B \rar \rho}_1(0) \simeq -10.5 \ \mbox{GeV}^{-2}
 \sqrt{M_b}\frac{<\bar dd>(\tau')}{\ga E^B_c\dr^{3/2}}
  \aga 1 +\frac{2.5\pm 1.1}{M_b} +
\frac{6.3\pm 1.1}{M^2_b}\adr,
\eeq
where each coefficient compares reasonably well with that of the
expanded expression. We have absorbed the error due to $E^B_c$ into
the errors in the corrections. One should understand in the previous
formula that:

\nin
The overall factor 10.5 is fixed in such a way that
 the interpolating formula reproduces, with the
central values of the numbers in (22), the numerical estimate in (16).
This factor also
absorbs in it the effect of the mixed condensate
$M^2_0$ as given by
 $\delta^{(0)}$. However,
if we assume that the factorisation works for the high-dimension
condensates (however, it is known to be largely violated by a factor 2
to 3
\cite{SN3}), we
could resum all light-quark-like condensates effects ( idea behind the
notion of non-local condensate)
by replacing $<\bar dd>$ with $<\bar dd>$exp$(-M^2_0\tau'/3)$, which
converges perfectly for $M_b \rar \infty$ contrary to the case of
the DLSR mentioned by \cite{ALI}.

\nin
The $1/M_b$ correction is mainly due to
$f_B$ (one should compare this coefficient with the one of
$f_B$ including the $1/M^2_b$ term (see e.g. \cite{SN1}))and to the
meson-quark mass-difference $\delta M_{(b)}$ (see (20)).

\nin
The
 coefficient of the $1/M_b^2$ term comes,
 partly from an $1/M_b$-expansion of $f_B$, which is known
to give  a quite good
approximation of $f_B$
even at the $c$ quark mass. However, the main contribution
to the $1/M^2_b$ term in (22) comes from the perturbative
vertex diagram. One should understand that the extraction of
this effect comes from the
exact non-expanded expression of the Wilson coefficient {\it
without any
 approximation} related to
the large value of $M_b$. The appearance of the $1/M^2_b$ term is only
due to the analytical structure of the perturbative contribution, and
its numerical coefficient
absorbs in it all the effects of the perturbative graph. From this
feature, the dominance of this term at the $c$ quark mass does not mean
that the formula given in (22) cannot be used at this scale. One should
understand (22) as an $interpolating$ formula.

\nin
This sum rule explicitly indicates that the $M_b$ behaviour is dominated
by the soft process term $<\bar qq>$ instead of
the perturbative
hard diagram for $M_b \rar \infty$. This is a peculiar feature of
this heavy-to-light transition process, which is not the case of the
heavy-to-heavy one.
We also obtain a similar behaviour in the case of the semileptonic
$B \rar \pi \ e\nu$ form factor \cite{SNP}.
However,
the authors of Refs.\cite{ALI}, \cite{RUC}
who work with the light-cone sum rule
do not have this dominant behaviour in $M_b$,
since
in their case the behaviour $M_b^{-3/2}$ is similar to that coming
from the perturbative graph, which is non-leading in $M_b$,
in our approach based on the SVZ-expansion.
A clarification of this disrepancy needs a better understanding of
the structure of the meson wave functions used in this analysis for
$M_b \rar \infty$.
The $M_b$ dependence obtained here at $q^2=0$ is the same as the one
obtained by Isgur-Wise at $q^2_{max}$, which might be in line with the
Isgur-Wise conjecture that the relations among form factors can be valid
at any $q^2$ values
if the heavy-to-light transition
form factors are dominated by the soft process: in such a case,
the heavy quark stays almost on its mass shell.
The dominance of the soft process obtained by Burdman-Donoghue \cite{DON}
is confirmed, in a completely independent way, by our analysis.

\nin
However, at the real value of $M_b$,
the agreement of the different previous sum-rule
predictions for the $B$ and $D$ meson form factors is encouraging,
despite the fact
that the sum rule used in \cite{COL} is not well-defined for $M_b \rar
\infty $ (however, truncating their QCD series at the level of the quark
condensate, one can notice that their result
 has the same $M_b$-behaviour than
ours (see their equation (24)),
while the results
of \cite{ALI} have a similar
$M_b$-behaviour than our non-leading perturbative contribution.
The  agreement between different numerical
estimates might be due to the large
numerical value of the $M_b^{-3/2}$-term, which can
also
invalidate the na\"\i ve extrapolation of the result from the
$D$ to the $B$ when only the leading $M_b$ behaviour of the
form factor is used.
 Using the previous interpolating formula at the $D$ mass,
we obtain:
\beq
F^{D\rar \rho}_1(0) \simeq (61.5\pm 30.5)\times 10^{-2},
\eeq
in accordance with the previous result in (17) from a numerical fit.
\section{\boldmath{$SU(3)_f$}-breakings and the
\boldmath{$B \rar K^* \ \gamma$} process }
We shall consider in this section the explicit $SU(3)_f$-breakings
on the form factor of the $B\rar K^* \ \gamma$, process,
due to the $s$ quark mass and to the $<\bar ss>$ condensate, which
have the values \cite{SN3, SN5}:
\beq
\bar {m}_s (\tau') \simeq 150 \; \mbox{MeV} \; \; \; \; \; \; \; \; \; \;
\frac{<\bar ss>}{<\bar dd>}
\simeq (0.6 \pm 0.1).
\eeq
For this process, the QCD expression of the $SU(3)_f$ breaking parts
of the vertex function reads to
leading order in $\alpha_s$:
\bea
\mbox{Im} V(s,s')_{SU(3)}&\rar&\frac{3}{8} M^2_bm_s \frac{s-s'-M^2_b}
{(s-s')^2} \nnb \\
 <\bar qq>_{SU(3)}&\rar& -\frac{m_s}{2}M_b<\bar dd>.
\eea
One can notice that the only effect of the quark condensate
which survives after the sum rule
procedure is the one from the $d$ quark line of the $B$-meson current,
such that only the $<\bar dd>$ condensate contributes.
Using an analytical approximate evaluation of the $SU(3)_f$-breaking
 effects, we deduce
the $K^*$ version of the interpolating formula in (22):
\bea
F^{B \rar K^*}_1(0) &\simeq& -11.3 \ \mbox{GeV}^{-2}
 \sqrt{M_b}\frac{<\bar dd>(\tau')}{\ga E^B_c\dr^{3/2}} \nnb \\
& &  \aga 1 + \frac{\bar {m}_s}{M_b}+\frac{2.5\pm 1.0}{M_b} +
\frac{6.3\pm 1.1}{M^2_b}\ga 1 +2\frac{\bar {m}_sE^B_c}{t'_c} \dr\adr,
\eea
where we have used $\gamma_{K^*} \simeq (2.80 \pm 0.13)$ from $\tau$
decay.
One should notice that the $SU(3)_f$-breakings due to the meson
mass and coupling give an effect of +7.7 \%, which is included in the
overall factor.
 The explicit breaking
due to the $s$ quark mass from the Wilson coefficient
of the condensate  contributes as +3.0\%.
The $SU(3)$ breaking from the perturbative diagram is about
23\% of the perturbative contribution and might explain the large
$SU(3)_f$-breakings obtained in \cite{ALI}.

\nin
One can also
note that the $SU(3)_f$-breakings vanish for $M_b \rar \infty$, contrary
to the case of $f_{B_s}$ where these corrections
remain constant \cite{SNS}.
We deduce from the previous formula:
\beq
{F^{B\rar K^*}_1}(0)/
{F^{B\rar \rho}_1}(0) \simeq  (1.14 \pm 0.02), \; \; \; \;
{F^{D\rar K^*}_1}(0)/
{F^{D\rar \rho}_1}(0) \simeq  (1.22 \pm 0.04),
\eeq
where it is clear that the systematic errors in the evaluation of the
coefficients of the $1/M_b$ and $1/M^2_b$ cancel out in the ratio. The
quoted error in (27) is mainly due to $SU(3)_f$-breaking from the
perturbative graph.
The $SU(3)_f$-breaking corrections are smaller than the ones
in Ref. \cite{ALI} as explained before. Combining this result with the
one in (16), we deduce:
\beq
{F^{B\rar K^*}_1} \simeq (30.8 \pm 1.3 \pm 3.6 \pm 0.6 ) \times 10^{-2},
\eeq
where the last error is due to (27).
This implies the branching
ratio:
\beq
Br(B\rar K^*\ \gamma) \simeq (4.45\pm 1.12)\times 10^{-5},
\eeq
in agreement with the CLEO data of $(4.5 \pm 1.5 \pm 0.9)\times 10^{-5}$
\cite {CLEO}. Our result for  $F^{B\rar K^*}_1(0)$
is in agreement with the one obtained from an effective lagrangian
approach \cite{GAT}. It also agrees with the value
$(0.32 \pm 0.05)$ obtained from the light-cone sum rule \cite{ALI},
which, {\it a priori}, is an approach quite different from ours, as
indeed, the two approaches do not provide the same large $M_b$-behaviour
of the form factor.
A comparison with the value
$(0.35 \pm 0.05)$
obtained in \cite{COL} from the DLSR is not very informative as this
 number comes for the uses of
inconsistent sets of input parameters as noticed by \cite{ALI} (we agree
with these criticisms). Moreover,
the analysis of \cite{COL} also suffers from
the bad
behaviour of the DLSR which blows up for large $M_b$.

\section{Conclusions}
We have estimated the $B \rar \rho \  \gamma$ form factor
in (16). The value of the hypothetical $D \rar \rho  \  \gamma$
form factor is given in (17). The value of the $B \rar K^* \ \gamma$
form factor including $SU(3)_f$-breakings is given in (28) and leads
to the branching ratio in (29).

\nin
 We have
performed
our analysis with the so-called {\it hybrid} sum rule (HSR)
in (6), which is
well-defined for $M_b \rar \infty$, contrary to the case of the double
exponential Laplace sum rule (DLSR) which blows up in this peculiar
process. Our numerical estimates are
in agreement with the previous results from light-cone
sum rule \cite{ALI}, though the two results do not have the same
large $M_b$-behaviour.

\nin
Indeed,
we have also studied the $M_b$-dependence of the previous
transition form factor which can be parametrized with the interpolating
formula in (24). It explicitly shows
that, in the large-mass limit, this form factor is
dominated by the light quark condensate and behaves like $\sqrt{M_b}$,
though at low $M_b$ mass
the perturbative contribution is $numerically$ important.
This dominance of the soft $<\bar dd>$ contribution, which is in line
with the results of Isgur-Wise \cite{IWI2}
and Donoghue-Burdman \cite{DON}, might allow the
extension, of the result obtained at $q^2_{max}$, to the whole range
of $q^2$-values.
In this large mass limit, the light-cone sum rule result has a
$M_b^{-3/2}$ behaviour, which is very similar to the one
from the perturbative diagram, in our approach
within the SVZ-expansion. A clarification of this discrepancy between
our result with the one from the light-cone, needs a
better understanding of the meaning of the $<\bar qq>$ condensate
in the language of the wave functions.

\nin
Finally, we have extracted an analytic expression of the
$SU(3)_f$-breaking terms due to the $s$ quark mass in (26) and (27).
 It shows, that the
$SU(3)_f$-breakings tend to zero for $M_b \rar \infty$ in contrast with
the case of $f_{B_s}$\cite{SNS}. Our numerical value in (27)
is smaller than
the one in \cite{ALI}.

\section*{Acknowledgements}
I have
enjoyed useful conversations with Ahmed Ali, Vladimir Braun and Antonio
Masiero.
\noindent
\section*{Figure captions}
$\tau'$ and n dependences of the form factor $F^{B\rar \rho}_1(0) $
of the $B \rar \rho \ \gamma$ process.
\vfill\eject

\end{document}